\documentclass[reprint,amsmath,amssymb,aps,nofootinbib,prd]{revtex4-1}

\usepackage[english]{babel}
\usepackage[utf8]{inputenc}
\usepackage[T1]{fontenc} 

\usepackage{amsmath,amssymb}
\usepackage{amscd}
\usepackage{amsfonts,palatino}
\usepackage{textcomp}
\usepackage{slashed}
\usepackage{cancel}

\usepackage{pb-diagram}
\usepackage{graphicx}
\usepackage{epstopdf}
\usepackage{color}
\usepackage[usenames,dvipsnames,svgnames,table]{xcolor}

\usepackage[bookmarks=false,urlcolor=NavyBlue,colorlinks=true,linkcolor={NavyBlue},citecolor=NavyBlue,pdfstartview={XYZ null null 1.22},backref,pagebackref]{hyperref}

\usepackage{paralist}
\usepackage{dcolumn}

\usepackage{bm}

\usepackage{layout}

\renewcommand{\cal}[1]{\mathcal{#1}}
\DeclareGraphicsExtensions{.pdf,.png,.jpg,.eps}

\begin{document}

\title{Electric-magnetic duality in linearized Ho\v{r}ava-Lifshitz gravity}
\author{I. Cortese}
\email{icortese@ulb.ac.be}
\affiliation{Physique Th\'eorique et Math\'ematique \& International Solvay Institutes\\
Universit\'e Libre de Bruxelles,\\
Campus Plaine C.P. 231, B-1050 Bruxelles, Belgique}

\author{J. Antonio Garc\'ia}
\email{garcia@nucleares.unam.mx}
\affiliation{Departamento de F\'isica de Altas Energ\'ias, Instituto de Ciencias Nucleares, \\ Universidad Nacional Aut\'onoma de M\'exico,\\
Apartado Postal 70-543, M\'exico D.F. 04510, M\'exico}

\begin{abstract}
Known as a symmetry of vacuum Maxwell equations, the electric-magnetic duality can be lifted actually to a symmetry of an action. The Lagrangian of this action is written in terms of two vector potentials, one electric and one magnetic, and while it is manifestly invariant under duality rotations, it is not manifestly Lorentz covariant. This duality symmetry exists also in linearized gravity in four dimensions, and can be lifted off shell too. In $d$ dimensions, the link between linearized gravity and its dual can also be seen from the point of view of a parental action. This is defined by a first order Lagrangian (with the help of some auxiliary variables) that delivers both Fierz-Pauli theory and its dual. In this work we use this formalism to implement the electric-magnetic duality in the nonrelativistic deviation of Fierz-Pauli theory arising from Ho\v{r}ava-Lifshitz gravity. Because this theory breaks diffeomorphism invariance, one finds that such implementation includes some peculiarities.
\begin{description}
\item[PACS numbers] 04.20.Fy, 04.50.Kd
\end{description}
\end{abstract}

% \pacs{04.20.Fy, 04.50.Kd}

\maketitle

%%%%%%%%%%%%%%%%%%%%%%%%%%%%%%%%%%%%%%%%%%%%%%%%%%%%%%%%%%%%%%%%%%%%%%%%%%%%%%%
%%%%%%%%%%%%%%%%%%%%%%%%%%%%%%%%%%%%%%%%%%%%%%%%%%%%%%%%%%%%%%%%%%%%%%%%%%%%%%%

%%%----------------------------------------------------------------------------
\section{Introduction \label{sec:intro}}
%%%----------------------------------------------------------------------------

Variational principles and symmetries are fundamental tools for the construction of interactions and the study of possible extensions of well-known classical field theories like gravity. Imposing global and local symmetries, together with unitarity and locality, provides us with a guide to explore extensions (or modifications) of the action of a given consistent theory.

What about duality symmetries? Particularly, what about the electric-magnetic duality invariance? This symmetry shows up in diverse physical systems, from zero to higher spin theories, and carries many interesting features that make it depart somehow from the general character of global and local symmetries. In the case of linearized gravity and free higher spin gauge fields in four dimensions, for example, duality invariance can be manifestly implemented at the level of the action and not just in the equations of motion. Such implementation breaks, however, manifest spacetime covariance \cite{DeTe,HeTe2,DeSe2}. This feature bookmarks an interesting link between duality and spacetime symmetries, something that acquires a concrete form in the case of an Abelian gauge field for which it can be shown that duality invariance implies Poincaré invariance \cite{BuHe4}. Moreover, evidence on the relation between duality and supersymmetry has been provided recently, showing that self-duality may arise in some models with $N=2$ supersymmetry plus a particular hidden $N=2$ supersymmetry (see \cite{CaKal}). On the other hand, an actual breaking of gauge invariance in Maxwell theory results if one tries to make duality invariance a local symmetry \cite{MaPr,Saa,BuHe3}. 

Another interesting feature of duality invariant models is that making it manifest at the level of the action requires new variables, ``electric'' and ``magnetic'' potentials, and doubling the number of gauge symmetries (see \cite{BaGo,BaTr2} for the case of electromagnetism and gravity with electric and magnetic sources). Despite these sorts of frictions between duality invariance and other symmetries in field theory, the interest in it comes from the belief that understanding an eventual relation between the two of them might enrich our view of hidden symmetries in supergravity and M theory \cite{West}. 

In $d$ dimensions, one has for gravity that the electric-magnetic duality relates the graviton, a rank-2 symmetric tensor, with fields of higher rank. If we account for duality, then there are two equivalent descriptions of linearized Einstein theory: One based on the metric $h_{\mu\nu}$ and the other in terms of its dual field $T_{\mu_1\ldots\mu_{d-3}|\nu}$, a mixed Young tableaux of $(d-3,1)$ symmetry type whose covariant action was given by Curtright \cite{Cu}. In four dimensions the dual graviton is a rank-2 symmetric tensor, as is the graviton itself. At nonlinear level, however, there are obstructions for the implementation of duality in the Einstein-Hilbert action. This is because the Curtright action does not admit any non-Abelian deformation under the assumptions of spacetime covariance and locality \cite{BeBoHe,BeBoCn}. 

The search for modifications and/or deformations of general relativity (GR) is actually a very active trend of current research. The aim there is to find a gravity theory that bypasses the basic problems associated with the quantization of the classical formulation of gravity as a standard field theory. Among the many directions in this trend one can find noncommutative gravity, stringy corrections to the GR action, new massive gravity, double field theory and, recently, Ho\v{r}ava-Lifshitz gravity \cite{Hor4,HoMe-Th,Hor3,Hor2,Hor1}.  The latter is proposed as a UV completion of GR using a ``heretic'' proposal \cite{KeSf}: breaking diffeomorphism invariance to a subgroup that is still manifest in ADM formalism. As a consequence of this breaking, different treatments for the kinetic and potential terms are necessary. The kinetic term is still quadratic in time derivatives (but deformed by introducing a new parameter $\lambda$), and the potential has terms that can be higher order in space derivatives of the metric. 

The Ho\v{r}ava-Lifshitz (HL) action
\begin{multline*}
S_{\text{{\scriptsize HL}}}=\int \mathrm{d}t\mathrm{d}^3x\sqrt{g}N\Bigl( \frac{2}{\kappa^2}( K_{ij}K^{ij}-\lambda K^2 ) +\mu^4 \stackrel{(3)}{R} \\+\zeta \stackrel{(3)}{R^2} +\chi \stackrel{(3)}{R^{ij}} \stackrel{(3)}{R}_{ij}+\gamma \stackrel{(3)}{R}\Delta \stackrel{(3)}{R}+\dots \Bigr),
\end{multline*}
where $K_{ij}$ and $\stackrel{(3)}{R}_{ij}$ denote, respectively, the components of the extrinsic curvature and the Ricci tensor of the spatial manifold, is built to be power counting renormalizable. At large distances, higher derivative terms do not contribute and the theory is supposed to run to GR. The parameter $\lambda$ is the parameter which controls the contribution of the trace of the extrinsic curvature and of some of the higher derivative terms.\footnote{Like the term proportional to $\stackrel{(3)}{R^2}$; for example, see \cite{Hor2}.} For $\lambda=1$ the theory has a fixed point in the IR, recovering GR. 

The aim of the present note is to implement the electric-magnetic duality in the linearized part of the HL action. The interesting point is that the breaking of full diffeomorphism invariance for generic values of $\lambda$ is not an obstacle for the implementation of duality. Then, this can be viewed as an exercise to test how duality is affected when we put spacetime covariance under stress.

The implementation of duality as a symmetry of the action of linearized gravity was done in \cite{HeTe2}. The approach of the authors in that work is to solve the Hamiltonian and momentum constraints to obtain the prepotentials (analogous to the magnetic and the electric vector potentials in electromagnetism ---the latter arising from solving the Gauss constraint), and then to show the manifest invariance of the action under $SO(2)$ rotations of those variables. Here we do not follow this method but instead we parallel the work of \cite{BoCnHe} where the authors connect Fierz-Pauli (FP) theory with its dual through a parental action; still, we do make some comments on the former approach in Sec. \ref{sec:diracanalisis}, where we fix some ideas related to the model at hand.

The parent action approach is defined with a first order Lagrangian from which both FP theory for the graviton in the framelike formulation and the corresponding theory for the dual graviton can be extracted (see Fig. \ref{fig:parentalscheme}). Duality in the framelike formulation has been worked out also in \cite{AjHaSi}. Having this scheme in mind, we proceed in Sec. \ref{sec:dualHL} in two steps: First we look for an action in the \textit{vielbein} basis for the linearized part of the HL gravity in Sec. \ref{sec:linHLvierbein}. This action is enhanced with respect to the latter in the sense that it enjoys diffeomorphism invariance, but still is not local Lorentz invariant; linearized HL gravity arises from it by imposing a gauge fixing condition. Then we find a parent action in Sec. \ref{sec:HLparental} that delivers this theory and its dual by the same mechanism used in \cite{BoCnHe}.
\renewcommand{\dgeverylabel}{\displaystyle}
\begin{figure}[ht!]
$$
\begin{diagram}
\node{}\node{\mbox{\small{Parental}}~\cal{L}[e,Y]}\arrow{se,t}{e_{ab}~\mbox{\small{multiplier}}}\arrow{sw,t}{Y^{ab\mid}_{}{c}~\mbox{\small{auxiliary}}}\node{} \\
\node{\boxed{\cal{L}[e_{ab}]}}\arrow{s,l}{\mbox{\small{gf}}~e_{[ab]}}%\arrow{ne,b}{}
\node{}\node{\boxed{\cal{L}[W^{abc\mid}{}_{d}]}}\arrow{s,l}{\mbox{\small{gf}}~W^{abc\mid}{}_{c}}\\
\node{\cal{L}_{\text{{\scriptsize FP}}}[h_{ab}]}%\arrow{n,r}{}
\node{}\node{\mbox{\small{dual of FP:}}~\cal{L}_{\text{{\scriptsize C}}}[T_{a_1\dots a_{d-3}\mid b}]}
\end{diagram}
$$
\caption{\textsf{\small{Role of the parent Lagrangian in $d$ dimensions for FP theory. $h_{ab}$ are the first order perturbations of the metric and $T_{a_1\dots a_{d-3}\mid b}$ are the components of the dual graviton. The child theories $\cal{L}[e_{ab}]$ and $\cal{L}[W^{abc\mid}{}_{d}]$ depend on the \textit{vielbein} and some potentials $W^{abc\mid}{}_{d}$, respectively (such that $Y^{ab\mid}{}_c=\partial_d W^{abd\mid}{}_c$, $Y^{ab\mid}{}_c$ being a redefinition of the spin connection). They enjoy arbitrary shifts of $e_{[ab]}$ and of the trace $W^{abc\mid}{}_{c}$ as gauge symmetries, respectively. Those variables can be marginalized with gauge fixing (gf) conditions recovering the FP Lagrangian on the left branch and the Curtright theory on the right one. In $d=4$ the dual graviton is a rank-2 symmetric tensor and $\cal{L}_{\text{{\scriptsize C}}}$ has the same functional form as $\cal{L}_{\text{{\scriptsize FP}}}$, showing that FP theory is self-dual.}}}\label{fig:parentalscheme}
\end{figure}

The output of the procedure described above in the case of $d=4$ is that the dual of linearized HL gravity is the same functional deviation of the FP Lagrangian for the dual graviton, but controlled by a different parameter $\gamma=\gamma(\lambda)$. In Sec. \ref{sec:Curtright}, and before some concluding remarks, we speculate in a sort of Ho\v{r}ava-like deviation of the Curtright action, the dual of FP theory in $d=5$.

%%%----------------------------------------------------------------------------
\section{Linearized minimal deviation\label{sec:diracanalisis}}
%%%----------------------------------------------------------------------------

In this section we review some properties of the model under consideration, its gauge symmetries and degrees of freedom (d.o.f.). Particularly we want to comment on the constraint structure of linearized HL gravity, and in its regards to the electric-magnetic duality under the light of the analysis performed in \cite{HeTe2}.

HL gravity comes with a preferred foliation of spacetime arising from an assumed inherent anisotropy scaling between space and time. In contrast to GR, the gauge symmetries of this theory are given by the \textit{foliation preserving diffeomorphisms}, generated by $\delta x^i=\xi^i(x,t)$ and $\delta t=\xi^{0}(t)$. Time reparametrizations $\delta t=\xi^{0}(t)$ are time dependent only and it has been shown that they are trivial gauge symmetries \cite{HeKlLu,HeKlLu2}. We will come back to this feature in the next section. 

As a consequence of the diffeomorphism invariance breaking, HL gravity contains a problematic extra mode \cite{ChNiPaSa}, which in the linear approximation is manifest only for perturbations around spatially inhomogeneous and time-dependent backgrounds \cite{BlPuSi2}. The theory generically exhibits also a dynamical inconsistency: In asymptotically flat spacetimes, the lapse function $N$ is generically constrained to vanish, preventing any interesting dynamics except for some particular cases (like spacetimes with vanishing extrinsic curvature) \cite{HeKlLu}. Despite these unappealing features, HL theory provides us with a concrete model for gravity that departs from the relativistic paradigm and is tractable for testing the electric-magnetic duality in it.
 
For the present analysis we assume the theory is describing some solutions that are blind to the above mentioned problems, and we take the linearized part of the minimal deviation from GR\footnote{The minimal deviation has two parameters besides the coupling $\kappa$ of GR. These are the dimensionless $\lambda$ in the kinetic term and $\mu$ with mass dimensions $[\mu]=1$ for the potential energy. We choose here $\frac{\mu^4}{2}=\frac{1}{\kappa^2}$ just to deal with the deviation due to $\lambda$. This action reproduces the EH action of GR in the $\lambda\rightarrow 1$ limit (see \cite{KeSf}).} represented by the action 
\begin{equation*}
S_{\text{{\scriptsize HL}}}=\frac{2}{\kappa^{2}}\int \mathrm{d}t\mathrm{d}^3x\sqrt{g}N\Bigl( K_{ij}K^{ij}-\lambda K^2  +\stackrel{(3)}{R} \Bigr)
\end{equation*}
in $d=4$ dimensions and ADM separation. The perturbation around flat spacetime $g_{ij}=\eta_{ij}+\varepsilon h_{ij}$, $N=1+\varepsilon n$, and $N_i=\varepsilon n_i$, to quadratic order gives the Lagrangian
\begin{multline}
\label{eq:lagADMlin}
L^{(2)}_{\text{{\scriptsize HL}}}=\varepsilon^{2}\int \mathrm{d}^3x \Bigl[\frac12 \dot{h}_{ij}\dot{h}^{ij}-\frac{{\textcolor{Red}{\lambda}}}{2}\dot{h}^2-2\partial_i n_j(\dot{h}^{ij}-{\textcolor{Red}{\lambda}}\eta^{ij}\dot{h}) \\ +\partial_i n_j\partial^i n^j +(1-2\textcolor{Red}{\lambda})(\partial_i n^i)^2  +R ~\Bigr]
\end{multline}
($h\equiv \eta^{ij}h_{ij}$ and $\dot{}\equiv\frac{\partial}{\partial t}$), with
\begin{multline}
\label{eq:grpot}
R \equiv -\frac12 \partial_k h_{ij} \partial^k h^{ij} + \frac12 \partial_i h \partial^i h 
+\partial_i h^{ij} \partial^k h_{kj}\\ 
- \partial_i h^{ij} \partial_j h +2n(\partial_i\partial_j h^{ij}-\Delta h).
\end{multline}
From now on we will focus our attention on the structural properties of Lagrangian (\ref{eq:lagADMlin}). This Lagrangian deviates from FP theory in those terms proportional to $\lambda$, to which this theory goes in the $\lambda\to 1$ limit. Despite that such modifications due to a value of $\lambda\neq 1$ are subtle, the structural consequences are big.

Foliation preserving diffeomorphisms act on the field variables as
\begin{equation}
\label{eq:folpresdiff}
    \delta h_{ij}=\partial_{i}\xi_{j}+\partial_{j}\xi_{i}, \quad\delta n_i= \dot{\xi}_i,\quad\delta n=-\dot{\xi}_{0},
\end{equation}
and they are a subset of the full group of diffeomorphisms: $\delta_{\xi}h_{\mu\nu}=\partial_{\mu}\xi_{\nu}+\partial_{\nu}\xi_{\mu}$. Having less gauge symmetry than the FP Lagrangian, we would expect linearized HL gravity to have more than 2 d.o.f.. However, this is not seen by the constraint structure of the model. We briefly summarize its analysis here because 
it shows a detail that we use in the argument of the next section.

The highlights of the constraint analysis go as follows: We get from the start the set of primary constraints $\phi^{0}=p$ and $\phi^{i}=p^{i}$, the variables $p,~p^{i}$ being the conjugate momenta of $n,~n_{i}$, respectively. The canonical Hamiltonian is given by
\begin{equation*}
H=\int \mathrm{d}^3x \Bigl( \frac12\pi^{ij}\pi_{ij}-\frac{\theta}{4}\pi^2+2\partial_i n_j\pi^{ij}-R\Bigr),
\end{equation*}
with
\begin{equation*}
\theta\equiv\frac{2\lambda}{3\lambda-1}, 
\end{equation*}
the variables $\pi^{ij}$ being the conjugate momenta of $h_{ij}$. The trace $\pi\equiv\eta_{ij}\pi^{ij}$ is proportional to the linearized trace of the extrinsic curvature:
\begin{equation*}
    \pi=(1-3\lambda)(\dot{h}-2\partial^in_i)=2(1-3\lambda)K.
\end{equation*}
For $\lambda=\frac13$ there is one more primary constraint: $\pi=0$. This case has been analyzed by the authors of \cite{BeReSo}, where they show that the same is equivalent to FP theory (and manifestly in the transverse-traceless gauge). In the following we exclude from our analysis this particular value of $\lambda$.

Stabilization of the primary constraints results in the set of secondary constraints,
\begin{equation*}
\psi^0=2(\partial_i\partial_j h^{ij}-\Delta h)\quad\text{and}\quad \psi^i=2\partial_j \pi^{ji}
\end{equation*}
(the same Hamiltonian and momentum constraints as in FP theory), and there are more secondary constraints:
\begin{equation}
\label{eq:K}
\chi=(\theta-1)\Delta\pi \quad \text{and} \quad \chi'=-4(\theta-1)\Delta^{2} n.
\end{equation}
$\chi=0$ results from stabilization of $\psi^{0}$, placing a constraint on the trace of the extrinsic curvature, and $\chi'=0$ from stabilization of $\chi$. At the end, stabilization of $\chi'$ results in conditions for a Lagrange multiplier ($\psi^{i}$ gives no further constraints nor conditions on Lagrange multipliers). Following the authors of \cite{HeKlLu}, the constraint equation $\Delta^{2}n=0$ in asymptotically flat spacetime determines the lapse function and in fact forces it to vanish, freezing in this way the evolution in time of the theory. The authors also showed (see \cite{HeKlLu2}) that time reparametrizations are trivial gauge transformations and so there is no first class constraint associated to it. Indeed, by inspection of the Poisson brackets between the constraints we see that they vanish except for
\begin{align*}
\{\chi(x),\psi^0(y)\}& = 2(\theta-1)\Delta^{2}\delta(x-y), \\
\{\chi'(x),\phi^0 (y)\}& = -4(\theta-1)\Delta^{2}\delta(x-y),
\end{align*}
and as a consequence the Hamiltonian constraint, the one we would have expected to generate the evolution in $t$, is second class. Thus we have a set of four second class ($\phi^{0},~\psi^{0},~\chi,~\chi'$) and a set of six first class ($\phi^{i},~\psi^{i}$) constraints, and thereby we might conclude that there are 2 d.o.f.. The fact that the extra mode is not manifest here is because we are taking the linearized theory around flat spacetime \cite{BlPuSi2}. Given that $\psi^{i}$ are first class, the only\footnote{Gauge transformations generated by constraints $\phi^i$ are just arbitrary shifts of $n_i$.} nontrivial gauge transformations of the Lagrangian (\ref{eq:lagADMlin}) are those associated to $\delta x^{i}=\xi^{i}(x,t)$.

Now, there is one detail of the constraint structure that is relevant for the electric-magnetic duality on Lagrangian (\ref{eq:lagADMlin}). Following the approach of \cite{HeTe2} for the case of FP, we can try to analyze duality rotations in terms of the prepotentials $\Phi_{ij}$ and $P_{ij}$ defined by solving the constraints $\psi^{0}=0$ and $\psi^{i}=0$ [see Eqs. (2.7) and (2.11) of \cite{HeTe2}]. However, in the present case we face an extra constraint placing a condition on $P_{ij}$, namely,
\begin{equation*}
\chi=(\theta-1)\Delta \pi=(\theta-1)\Delta (\eta^{ij}\Delta -\partial^{i}\partial^{j})P_{ij}=0,
\end{equation*}
with no counterpart for the $\Phi_{ij}$ variables. Besides, there is no exchange between the kinetic and potential energies after these rotations as it happens in FP theory, and the condition above does not help to make the match. This leaves us with the idea that linearized HL gravity is not self-dual. 

Nevertheless, we can analyze duality from another point of view that allows us to draw a map to the magnetic dual of linearized HL gravity ---we call the Lagrangian (\ref{eq:lagADMlin}) the electric one. The path goes through a parent action, analogous to the one constructed in \cite{CaMoUr3} for FP theory. This procedure is reformulated in \cite{BoCnHe} in a way that suits the purpose of the present study. In the following section, we use this approach to find the magnetic dual of linearized HL gravity.

%%%----------------------------------------------------------------------------
\section{Parental action and dual theory\label{sec:dualHL}}
%%%----------------------------------------------------------------------------
 
Schematically, the road map presented in \cite{BoCnHe} starts from the second order action in \textit{vielbein} formalism for FP, taking it up to a first order parent action and then down from it to the second order dual theory given by the Curtright Lagrangian (see Fig. \ref{fig:parentalscheme}). In the present case, and to keep parallelism with FP theory, we need an action in the \textit{vielbein} basis as a starting point, the same that should be obtained by a parental one (see left branch of Fig. \ref{fig:approach}). 

We proceed then in two steps. First we look for an action corresponding to the Lagrangian (\ref{eq:lagADMlin}) in the \textit{vielbein} basis in Sec. \ref{sec:linHLvierbein} and next we construct a parent action from which we can obtain the latter and its dual in Sec. \ref{sec:HLparental}. The Lagrangian in the \textit{vielbein} basis we write is in fact invariant under full diffeomorphisms, but not local Lorentz invariant though. It has more fields and symmetries than linearized HL gravity and delivers this theory through gauge fixing conditions. Finally, on the dual side, we define the theory, also imposing analogous gauge fixing conditions (see the right branch of Fig. \ref{fig:approach}).
\renewcommand{\dgeverylabel}{\displaystyle}
\begin{figure}[ht!]
$$
\begin{diagram}
\node{}\node{\mbox{\small{Parental}}~\cal{L}_{\lambda}[e,Y]}\arrow{se,t}{e_{ab}~\mbox{\small{multiplier}}}\arrow{sw,t}{Y^{ab\mid}{}_{c}~\mbox{\small{auxiliary}}}\node{} \\
\node{\boxed{\cal{L}_{\lambda}[e_{ab}]}}\arrow{s,l}{\mbox{\small{gf}}}%~\rho}%\arrow{ne,b}{}
\node{}\node{\boxed{\cal{L}_{\lambda}[W^{abc\mid}{}_{d}]}}\arrow{s,r}{\mbox{\small{gf}}}\\%~\rho*} \\
\node{\genfrac{}{}{0pt}{}{\underbrace{\cal{L}_{\text{{\scriptsize HL}}}[h_{ab}]}}{{\mbox{linear HL gravity}}}
%\cal{L}_{\text{{\scriptsize HL}}}[h_{ab}]
}%\arrow{n,r}{}
\node{}\node{\genfrac{}{}{0pt}{}{\underbrace{\cal{L}_{\text{{\scriptsize HL}}}[T_{a_1\dots a_{d-3}\mid b}]}}{{\mbox{dual of lin. HL gravity}}}
% \overbrace{\mbox{\small{dual of lin. HL grav:}}~\cal{L}_{\text{{\scriptsize HL}}}[T_{a_1\dots a_{d-3}\mid b}]}
}
\end{diagram}
$$
\caption{\textsf{\small{The Lagrangian density $\cal{L}_{\lambda}[e_{ab}]$ depends on the \textit{vielbein}s and it is parametrized by $\lambda$, the same parameter appearing in linearized HL gravity defined with Lagrangian (\ref{eq:lagADMlin}). This theory is diffeomorphism invariant, so it is not linearized HL gravity; the latter is obtained from $\cal{L}_{\lambda}[e_{ab}]$ by gauge fixing (gf) conditions that in fact eliminate all antisymmetric components of $e_{ab}$. Analogously on the dual side $\cal{L}_{\lambda}[W^{abc\mid}{}_{d}]$ has some variables that can be eliminated through gauge fixing conditions to get a Lagrangian depending only on the dual graviton components.}}}\label{fig:approach}
\end{figure}

\subsection{Linearized HL gravity in \textit{vielbein} language}
\label{sec:linHLvierbein}

The Lagrangian density 
\begin{equation*}
\cal{L}_{\text{{\scriptsize EH}}}[e_{ab}]=-\frac{e}{2}\Bigl( \Omega^{abc}\Omega_{abc}+2\Omega^{abc}\Omega_{acb}-4\Omega^{a}\Omega_{a} \Bigr)
\end{equation*}
represents Einstein-Hilbert (EH) theory in the \textit{vielbein} $e_{a}{}^{\mu}$ basis. The inverses of these fields are written\footnote{Latin letters from the beginning of the alphabet refer to indices in tangent space (contracted with the flat metric $\eta_{ab}$) whereas greek letters refer to indices in spacetime. Symmetrization and antisymmetrization, denoted by brackets $(\dots)$ and $[\dots]$, respectively, of any set of indices are defined with strength 1.} $e_{\mu}{}^{a}$, the functions $\Omega_{ab}{}^{c}\equiv 2 e_{a}{}^{\mu}e_{b}{}^{\nu}\partial_{[\mu}e_{\nu]}{}^{c}$ are the anholonomy coefficients, $\Omega_{a}\equiv\Omega_{ab}{}^{b}$ and $e\equiv \det{(e_{a}{}^{\mu})}$. This Lagrangian depends on $d^{2}$ fields in $d$ dimensions, certainly more variables than the independent components of the metric. However, local Lorentz invariance takes care of the excess.

Linearization of the above Lagrangian corresponds to FP theory, written as
\begin{equation}
\label{eq:FPlag}
\cal{L}_{1}[e_{ab}]=-\frac{1}{2}\Bigl( \Omega^{abc}\Omega_{abc}+2\Omega^{abc}\Omega_{acb}-4\Omega^{a}\Omega_{a} \Bigr),
\end{equation}
where $\Omega_{abc}=2\partial_{[a}e_{b]c}$. (We apologize for the abuse of notation when using the same for the \textit{vielbein} and its perturbation to linear order.) Two things are important to notice here: tangent and world space indices are identified, and the fields $e_{ab}$ do not have definite symmetry. Nevertheless, $\cal{L}_{1}[e_{ab}]$ depends on the symmetric part $h_{ab}\equiv 2e_{(ab)}$ and on the antisymmetric part $f_{ab}\equiv 2e_{[ab]}$ in very different ways. 

The above Lagrangian splits as
\begin{equation*}
    \cal{L}_{1}[e_{ab}]=\cal{L}_{\text{{\scriptsize FP}}}[h_{ab}] + \cal{M}[h,f],
\end{equation*} 
where 
\begin{multline*}
    \cal{L}_{\text{{\scriptsize FP}}}[h_{ab}]=-\frac12\partial^ah^{bc}\partial_ah_{bc}+\partial_ah^{ac}\partial^bh_{bc}\\
    -\partial^bh^{a}{}_{a}\partial^c h_{bc}+\frac12\partial_ah^{b}{}_{b}\partial^a h^{c}{}_{c}
\end{multline*}
is the FP Lagrangian for $h_{ab}$, and $\cal{M}$ is a total derivative, namely,
\begin{equation}
\label{eq:M}
\cal{M}\equiv \partial_{a}\Bigl(-\frac12 h^{c}{}_{c}\partial_{b}f^{ab}\Bigr).
\end{equation}
So, while $h_{ab}$ is a dynamical field,\footnote{In fact, $h_{ab}$ are the components in tangent space of the perturbation to the metric around flat spacetime: $h_{\mu\nu}=g_{\mu\nu}-\eta_{\mu\nu}$.} $f_{ab}$ enters the action only through a total derivative term, having relevance, if any, only for boundary conditions. 

Because of that, the Lagrangian density (\ref{eq:FPlag}) is invariant under the transformations $\delta e_{ab}=\partial_{(a}\xi_{b)}+\frac12\omega_{ab}$ given by general coordinate transformations $\delta x^{\mu}=\xi^{\mu}(x)$ and arbitrary shifts of the components $f_{ab}$ in terms of parameters $\omega_{ab}(x)=-\omega_{ba}(x)$. In other words, the Lagrangian for FP theory in this language is invariant under the field variations
\begin{equation}
\label{eq:FPgaugesymm}
\delta_\xi h_{ab}=2\partial_{(a}\xi_{b)}\quad\text{and}\quad\delta_\omega f_{ab}=\omega_{ab}.
\end{equation}
Using the gauge transformations for the components $f_{ab}$ we can fix them on the boundary to be zero, eliminating them from the theory (both from the action and boundary data). The Lagrangian (\ref{eq:FPlag}) is then a theory only for the symmetric components $h_{ab}$. We are looking for a Lagrangian density for linearized HL gravity just as a deviation from it. 

Let us think for the moment in $d=4$ and take the following prescription:
\begin{equation}
\label{eq:lagrhorlin}
    \cal{L}_{\lambda}[e_{ab}]=\cal{L}_{1}[e_{ab}]+2(1-\lambda)(\Omega_0)^2,
\end{equation}
where the parameter $\lambda$ is meant to be the same appearing in the Lagrangian (\ref{eq:lagADMlin}). Again, the above Lagrangian density depends on both the symmetric $h_{ab}$ and antisymmetric $f_{ab}$ parts of the \textit{vierbein}. One can anticipate that the presence of the last term for $\lambda\neq 1$ will break Lorentz invariance somehow and thus prevent allocating $f_{ab}$ inside a total derivative. These variables will be eliminated by slightly different means to get a theory just for the symmetric components $h_{ab}$, the variables which linearized HL gravity depends on.

To see that, let us inspect the symmetries of the Lagrangian (\ref{eq:lagrhorlin}), where 
\begin{equation*}
\Omega_0 = \frac12 [\dot{h} -\partial^i(h_{0i}+f_{0i})],\quad h\equiv\eta^{ij}h_{ij},
\end{equation*}
and which we do not expect to be more than those of Eq. (\ref{eq:FPgaugesymm}). Here we assumed some choice of reference frame where $x^{0}\equiv t$ is  the time coordinate. Both $\Omega_{0}$ and $\cal{L}_{1}$ depend on $h_{ab}$ and $f_{ab}$, but in the latter $f_{ab}$ enters only the total derivative $\cal{M}$ given in (\ref{eq:M}). For the following, it is convenient to make use of the ADM separation of variables, for which $n_i= h_{0i}$, and we define the notation $e_i\equiv f_{0i}$. 

The way in which the Lagrangian (\ref{eq:lagrhorlin}) is written reflects that it does not share all the symmetries of $\cal{L}_{1}$ given in (\ref{eq:FPgaugesymm}). Let us concentrate on how the $(\Omega_0)^2$ term changes under those transformations. Despite the manifest fact that it is not a Lorentz covariant quantity, it is invariant under the global variations $\delta_{\omega}f_{ab}$ with parameters $\omega_{ab}$ being constant.\footnote{In general $\delta\Omega_{abc}=\partial_{c}\partial_{[a}\xi_{b]}+\partial_{[a}\omega_{b]c}$, so for $\xi_a$ linear in $x^\mu$ and $\omega_{ab}$ a constant, we have $\delta\Omega_{abc}=0$.} The same can be cast as invariant under general coordinate transformations by breaking local Lorentz invariance. To see how, let us apply the transformations (\ref{eq:FPgaugesymm}) on it. That is,
\begin{multline*}
\delta (\Omega_0)^2=2\Omega_0\delta \Omega_0=\Omega_0\bigl(\delta_\xi\dot{h}-\partial^i(\delta_\xi n_i)-\partial^i(\delta_\omega e_i)\bigr) \\
=\Omega_0\bigl( 2\partial^i\dot{\xi}_i-\partial^i(\dot{\xi}_i+\partial_i\xi_0)-\partial^i\omega_{0i}\bigr) \\
=\Omega_0\bigl( \partial^i\dot{\xi}_i-\Delta\xi_0-\partial^i\omega_{0i}\bigr).
\end{multline*}
In order to have a symmetry, we have to require $\partial^i(\dot{\xi}_i-\partial_{i}\xi_0-\omega_{0i})=0$. We can take this as an equation for $\omega_{0i}$ restricting the arbitrariness of this parameter. The general solution for it is the sum of the solution to the homogeneous equation (any transverse $\zeta^{i}$) and a particular solution to the inhomogeneous equation:
\begin{equation}
\label{eq:omegarestr}
\omega_{0i}=\zeta_{i}+\dot{\xi}_{i}-\partial_{i}\xi_{0},\quad \text{such that} \quad\partial^{i}\zeta_{i}(x,t)=0.
\end{equation}

We observe that shifts of $f_{ij}$ with parameters $\omega_{ij}$ survive the deviation due to $\lambda\neq 1$ as gauge symmetries, and also the gauge transformations of the transverse part of $e_{i}$ with parameters $\zeta_{i}$. The transformation of the longitudinal part of $e_{i}$ on the other hand gets linked to the transformation of the longitudinal part of $n_{i}$, and hence their shifts are not arbitrary anymore. To sum up, the gauge symmetries of Lagrangian (\ref{eq:lagrhorlin}) are given by
\begin{subequations}
\label{eq:gaugsymm} 
\begin{align}
\delta_\xi h_{ij}=2\partial_{(i}\xi_{j)},\quad &\delta_\xi n_i=\dot{\xi}_i+\partial_i\xi_0,\quad \delta_\xi n= -\dot{\xi}_0, \label{eq:gaugsymmh} \\
\delta_{\omega} f_{ij}=\omega_{ij}, \quad &\delta_\omega e_i=\zeta_{i}+\dot{\xi}_i-\partial_i\xi_0. \label{eq:gaugsymmf} 
\end{align}
\end{subequations}
The transformations (\ref{eq:gaugsymmh}), which are just $\delta_{\xi}h_{ab}=2\partial_{(a}\xi_{b)}$ projected on ADM variables, show that $\cal{L}_{\lambda}$ enjoys full diffeomorphism invariance, and the (\ref{eq:gaugsymmf}) part that the variables $f_{ij}$ and the transverse part of $e_{i}$ are pure gauge (but not the longitudinal part of $e_{i}$, which is the one appearing in $\Omega_{0}$).

Let us now take a closer look into $\cal{L}_{\lambda}[e]$. By direct inspection we see that the set of equations of motion (EoM) of this Lagrangian for $h_{ij}, n_i, n, e_i$ is not equivalent to the one arising from linearized HL gravity. For instance, the EoM for $e_i$,
\begin{equation}
\label{eq:eome}
    -(1-\lambda)\partial_i(\dot{h}-\partial^jn_j-\partial^je_j)=0
\end{equation}
has no parallel in the set of EoM coming from Lagrangian (\ref{eq:lagADMlin}). Nevertheless, it is not difficult to see that if we enforce that $\partial^i n_i=\partial^i e_i$ in the first set, the two systems are equivalent. In the above equation, for example, what happens is that this condition is bringing the constraint on the linearized trace of the extrinsic curvature to a Lagrangian equation [{\em{cf.}} (\ref{eq:eome}] after this projection with constraint $\chi=0$ in (\ref{eq:K})).

The equivalence of these two systems can also be seen directly at the level of the Lagrangians. Using the shorthand $m_i\equiv n_i+e_i$ direct computation gives
\begin{align*}
\cal{L}_{\lambda}&= \cal{L}_{1}+\frac{(1-\lambda)}{2}(\dot{h} -\partial^i m_{i})^2 \\
&=\frac{1}{2}\Bigl(\dot{h}_{ij}\dot{h}^{ij}-\lambda \dot{h}^2 -4\partial_jn_i\dot{h}^{ij}\\ 
&\hskip0.3in+(4\partial^in_i -2\partial^im_i +2\lambda\partial^im_i)\dot{h}\\
&\hskip0.3in-2\partial^in_i\partial^jn_j+(1-\lambda)\partial^im_i\partial^jm_j+2\partial_in_j\partial^in^j \\
&\hskip0.3in-\partial_k h_{ij}\partial^k h^{ij} +\partial_i h \partial^i h +2\partial^ih_{ij}\partial_k h^{kj}\\
&\hskip0.3in- 2\partial_i h^{ij} \partial_j h +4n(\partial_i\partial_j h^{ij}-\Delta h)\Bigr) \\
&\hskip0.3in +\cal{M}[h_{ij},n_{i},n;f_{ij},e_{i}].
\end{align*}
It is straightforward to see that if we impose
\begin{equation}
\label{eq:HLgaugefixcond}
\rho\equiv\partial^i n_{i}-\partial^i e_{i}=0
\end{equation}
in the above expression, we have $\partial^im_i=2\partial^in_i$, and then
\begin{align*}
\cal{L}_{\lambda}\mid_{\rho=0}&=\frac{1}{2}\Bigl(\dot{h}_{ij}\dot{h}^{ij}-\lambda \dot{h}^2 -4\partial_in_j(\dot{h}^{ij} -\lambda\eta^{ij}\dot{h})\\
&\hskip0.3in+2(1-2\lambda)\partial^in_i\partial^jn_j+2\partial_in_j\partial^in^j \notag\\
&\hskip0.3in-\partial_k h_{ij}\partial^k h^{ij} +\partial_i h \partial^i h +2\partial^ih_{ij}\partial_k h^{kj} \\
&\hskip0.3in- 2\partial_i h^{ij} \partial_j h +4n(\partial_i\partial_j h^{ij}-\Delta h)\Bigr)\notag +\cal{M}\mid_{\rho=0} \notag \\
&=\cal{L}_{\text{{\scriptsize HL}}} +\cal{M}\mid_{\rho=0}.
\end{align*}

Hence, by imposing the condition (\ref{eq:HLgaugefixcond}) on $\cal{L}_\lambda[e_{ab}]$, we get the Lagrangian of linearized HL theory modulo the total derivative $\cal{M}\mid_{\rho=0}$, in which the longitudinal part of $e_{i}$ is fixed in terms of $n_{i}$. (That is all we can do using $\rho=0$, since this condition says nothing about the transverse part of this vector.) Now we can go further and get rid of $f_{ij}$ and the transverse part of $e_{i}$ inside $\cal{M}\mid_{\rho=0}$ by setting them to zero using the gauge symmetries given in (\ref{eq:gaugsymmf}).

In conclusion, taking $\cal{L}_{\lambda}$ simultaneously with $\rho=0$ we have a theory for the symmetric components $h_{ab}$ of the \textit{vierbein}s, whose dynamics is given precisely by the EoM of $\cal{L}_{\text{{\scriptsize HL}}}$ and for which we can provide boundary data that confidently allow us to forget all about $\cal{M}$. Moreover, the condition (\ref{eq:HLgaugefixcond}) is preserved by the gauge symmetries of linearized HL gravity. Its variation under transformations (\ref{eq:gaugsymm}) is
\begin{multline*}
\delta \rho=\partial^i(\delta_\xi n_i)-\partial^i(\delta_\omega e_i)\\
=\partial^i(\dot{\xi}_i+\partial_i\xi_0)-\partial^i(\dot{\xi}_i-\partial_i\xi_0)=2\Delta\xi_0.
\end{multline*}
Hence, $\rho=0$ is preserved by transformations with parameters $\xi^i=\xi^i(x,t)$ completely arbitrary and $\xi^{0}$ subject to equation $\Delta \xi_{0}=0$. Arbitrary functions of time $\xi^{0}=\xi^{0}(t)$ are solutions of this equation.

Actually, it is possible to see that $\rho=0$ is a partial gauge fixing condition that eliminates time reparametrizations as nontrivial gauge transformations. In the Appendix, \ref{app:newconstr} we perform the constraint analysis of the Lagrangian (\ref{eq:lagrhorlin}) and see that there are first class constraints generating, in particular, the gauge transformations
\begin{equation*}
\delta_\xi h_{ij}=0,\quad\delta_\xi n_i=\partial_i \xi_{0}=-\delta_\omega e_i,\quad \delta_\xi n= -\dot{\xi_{0}},
\end{equation*}
\textit{i.e.}, Eq. (\ref{eq:gaugsymm}) for $\xi^{\mu}=(\xi^{0},0,0,0)$ with $\xi^{0}=\xi^{0}(x,t)$ arbitrary. The condition (\ref{eq:HLgaugefixcond}) can be taken as a gauge fixing condition of one of those first class constraints. 

Now that we have a Lagrangian in the \textit{vierbein}s basis for linearized HL gravity (in the sense explained above and depicted in the left branch of Fig. \ref{fig:approach} we proceed next to find the parent action from which $\cal{L}_\lambda[e_{ab}]$ and its dual theory can be obtained.

\subsection{Parent action for linearized HL gravity and its dual}
\label{sec:HLparental}
 
We follow here the authors of \cite{BoCnHe}, where they define a parent action for FP theory,\footnote{We also use the notation of this reference.} to find the corresponding parent action for the deviation of this theory defined with Lagrangian (\ref{eq:lagrhorlin}). Roughly speaking, we look for a first order theory depending on the \textit{vielbein} and some auxiliary fields from which we extract both the second order electric Lagrangian $\cal{L}_{\lambda}[e_{ab}]$ and its magnetic dual. The former arises from solving the auxiliary fields into the parent action. The dual theory is obtained by taking $e_{ab}$ in it as Lagrange multipliers.

In addition to the procedure to obtain the dual of $\cal{L}_{\lambda}[e_{ab}]$, to actually get the dual of linearized HL gravity for the symmetric components of the \textit{vierbein}s in $d=4$, we have to impose a magnetic analogue of the electric condition (\ref{eq:HLgaugefixcond}).

We start by considering the Lagrangian (\ref{eq:lagrhorlin}),
\begin{equation*}
    \cal{L}_{\lambda}[e_{ab}]=\cal{L}_{1}[e_{ab}]+2(1-\lambda)(\Omega_0)^2 ,
\end{equation*}
in $d$ dimensions. A parent action for it can be defined with the Lagrangian density
\begin{multline}
\label{eq:lagrhorpo}
    \cal{L}_{\lambda}[e,Y]=- 2\Bigl( Y^{ab\mid c}\Omega_{abc}- Y^{ab\mid c}Y_{ac\mid b} +\frac{1}{(d-2)}Y^{a} Y_{a} \Bigr)\\
    +2\frac{(\theta -1)}{(d-2)}(Y_{0})^{2},
\end{multline}
with $\Omega_{abc}=2\partial_{[a}e_{b]c}$, and $\theta$ a parameter that we expect to be related with $\lambda$ (see below). Here, besides the \textit{vielbein} we have some tensor variables $Y^{ab\mid}{}_{c}$ antisymmetric in $ab$, but with no definite symmetry in $c$. As before, the notation $Y^{a}\equiv Y^{ab\mid}{}_{b}$ means the trace $Y^{ab\mid}{}_b=\eta_{bc}Y^{ab\mid c}$. 

We see from the EoM of the Lagrangian (\ref{eq:lagrhorpo}) that $Y^{ab\mid}{}_{c}$ are auxiliary fields, solved in terms of $e_{ab}$ and their derivatives as
\begin{equation}
\label{eq:yvarauxhor}
Y_{ab\mid c}=\frac12\Omega_{abc}-\Omega_{c[ab]}+2\eta_{c[a}\delta_{b]}{}^{d}\Bigl(\Omega_{d}-f(\theta)\eta_{d0}\Omega^{0}\Bigr), 
\end{equation}
with
\begin{equation*}
f(\theta)=\frac{(d-2)(\theta-1)}{(d-1)(\theta-1)+1}.
\end{equation*}
Plugging Eq. (\ref{eq:yvarauxhor}) into the first order Lagrangian (\ref{eq:lagrhorpo}) we get the second order electric $\cal{L}_{\lambda}[e_{ab}]$ if
\begin{equation*}
    \theta=\frac{(d-2)\lambda}{(d-1)\lambda -1}.
\end{equation*}
The parameter $\theta(\lambda)$ plays the role of $\lambda$ in the dual magnetic theory, a coefficient in front of the deviation from the dual of $\cal{L}_{1}[e_{ab}]$. Hence, the electric and magnetic theories are deviations from FP and its dual by terms proportional to parameters belonging to complementary regimes, since when $\lambda < 1 \Rightarrow \theta > 1$ and \textit{vice versa}.

Before going on to the dual theory, let us check the symmetry transformations of the parent Lagrangian. Transformations (\ref{eq:gaugsymm}) of $\cal{L}_{\lambda}[e_{ab}]$ can be extended to gauge invariances of $\cal{L}_{\lambda}[e,Y]$ by the corresponding transformations of the auxiliary fields. In fact, the restriction given in Eq. (\ref{eq:omegarestr}) arises here also from general assumptions. Using that $\delta e_{ab}=\partial_{(a}\xi_{b)}+\frac12\omega_{ab}~\Rightarrow~\delta\Omega_{abc}=\partial_{c}\partial_{[a}\xi_{b]}+\partial_{[a}\omega_{b]c}$, we get from Eq. (\ref{eq:yvarauxhor})
\begin{multline}
\label{eq:gaugtrY}
\delta Y_{ab\mid c}=\partial_{c}\partial_{[a}\xi_{b]}-\frac12\partial_{c}\omega_{ab}\\
+\eta_{c[a}\left(\eta_{b]d}-(1-\lambda)\eta_{b]0}\delta^{0}{}_{d}\right)(\partial^{d}\partial^{e}\xi_{e}-\partial^{e}\partial_{e}\xi^{d}+\partial_{e}\omega^{ed}).
\end{multline}
The Lagrangian is invariant,
\begin{multline*}
\delta \cal{L}_{\lambda}=-\Bigl( \delta Y^{ab\mid c}\Omega_{abc}+4\frac{(\theta-1)}{(d-2)}Y_{0}\delta Y_{0}\\
+Y^{ab\mid c}(\delta\Omega_{abc}-2\delta Y_{ac\mid b}+\frac{2}{(d-2)}\eta_{bc}\delta Y_{a})\Bigr)  \simeq 0,
\end{multline*}
where $\simeq$ means modulo total derivatives.

The first term in the last expression involves the $e_{ab}$ variables, and it is the only one that does. Requiring its invariance results in the differential equation that is solved with Eq. (\ref{eq:omegarestr}), restriction on the parameters that implies $\delta Y_{0}=0$, making the last term invariant on its own. The rest of the terms cancel out on account of the antisymmetry property of $Y^{ab\mid}{}_{c}$ in indices $ab$. Hence, the gauge symmetries of Lagrangian (\ref{eq:lagrhorlin}) are properly captured by symmetries of the parent Lagrangian.

We now compute the dual theory. Taking in the Lagrangian (\ref{eq:lagrhorpo}) the $e_{ab}$ as Lagrange multipliers, the constraints associated read $\partial_a {Y^{ab\mid}}_c=0$. These are solved by potentials ${W^{abe\mid}}_c$ completely antisymmetric in indices $abe$ such that ${Y^{ab\mid}}_c=\partial_{e}{W^{abe\mid}}_c$. Reduction of Lagrangian (\ref{eq:lagrhorpo}) gives the dual theory:
\begin{multline}
\label{eq:hordual1}
    \cal{L}_{\lambda}[{W^{abe\mid}}_{c}]=2 \Bigl(Y^{ab\mid c}Y_{ac\mid b}-\frac{1}{(d-2)}Y^{a}Y_{a}\Bigr)\\+2\frac{(\theta-1)}{(d-2)}(Y_{0})^{2},
\end{multline}
where $Y^{ab\mid}{}_{c}$ are functions of the potentials $W^{abe\mid}{}_{c}$. It is not difficult to see that the action defined with this Lagrangian is invariant under transformations (\ref{eq:gaugtrY}) for $Y^{ab\mid}{}_{c}$, the restriction (\ref{eq:omegarestr}) coming here from the last term.\footnote{With this restriction, using Eq. (\ref{eq:gaugtrY}) it can be seen that $\delta(\partial_{a}Y^{ab\mid}{}_{c})=0$, meaning that the constraint defining the dual potentials $Y^{abc\mid}{}_{d}$ is invariant for those gauge transformations.} However, these do not exhaust the gauge symmetries of the dual theory. There are more of these coming from an ambiguity in the potentials solving $\partial_a {Y^{ab\mid}}_c=0$. The redundancy is given by the gauge transformations
\begin{equation}
\label{eq:phiymm}
\delta_{\phi}W^{abc\mid}{}_{d}=\partial_{e}\phi^{abce\mid}{}_{d}
\end{equation}
with $\phi^{abce\mid}{}_{d}=\phi^{[abce]\mid}{}_{d}$ arbitrary. These transformations leave the field strengths $Y^{ab\mid}{}_{d}$ invariant and so the Lagrangian (\ref{eq:hordual1}).

A closer look into the symmetries of the dual magnetic theory and its structure comes from the decomposition
\begin{equation*}
{W^{abc\mid}}_d={X^{abc\mid}}_d+\delta_d^{[a}Z^{bc]}, 
\end{equation*}
consisting of extracting from ${W^{abc\mid}}_d$ its trace (the fields ${X^{abc\mid}}_d$ are completely antisymmetric in $abc$ indices and traceless). Plugging the above decomposition into Lagrangian (\ref{eq:hordual1}), we end up with the dual theory written as
\begin{equation}
\label{eq:hordual}
   \cal{L}_{\lambda}[X,Z]=\cal{L}_{1}[X,Z]+\frac{2(d-2)}{9}(\theta -1)(\partial_{i}Z^{i0})^{2}.
\end{equation}
The $\cal{L}_{1}[X,Z]$ part is the dual of the FP theory in $d$ dimensions:
\begin{equation*}
\cal{L}_{1}[X,Z]=2~\partial_{d}X^{abd\mid c}\partial^{e}X_{ace\mid b}+\mathcal{M}[X,Z],
\end{equation*}
the last term being the total derivative
\begin{equation}
\label{eq:totalderdual}
\cal{M}\equiv \frac23\partial_{c}\Bigl(2\partial_{d}X^{cda\mid b}Z_{ab}+ Z_{a}{}^{[b}\partial_{b}Z^{ac]}\Bigr)
\end{equation}
and the only place where $\cal{L}_{1}[X,Z]$ depends on $Z^{ab}$.

We identify Lagrangian (\ref{eq:hordual}) as the dual of (\ref{eq:lagrhorlin}). It is manifest here that $\cal{L}_{1}$ is invariant under arbitrary shifts of the traces: $\delta_{\tilde{\omega}}Z^{ab}=\tilde{\omega}^{ab}=-\tilde{\omega}^{ba}$. These include the projection of the gauge symmetries given in Eq. (\ref{eq:phiymm}) for those variables, which split as
\begin{multline*}
\delta_{\phi}W^{abc\mid}{}_{d}=\\
\partial_{e}\phi^{abce\mid}{}_{d}\rightarrow \left\{
\begin{aligned}
& \delta_{\phi}Z^{ab}= -\frac{3}{(d-2)}\partial_{c}\phi^{abc},~\phi^{abc}\equiv\phi^{abcd\mid}{}_{d}\\
& \delta_{\phi}X^{abc\mid}{}_{d} = \partial_{e}\phi^{abce\mid}{}_{d}+ \frac{3}{(d-2)}\delta_{d}^{[a}\partial_{e}\phi^{bc]e}
\end{aligned}
\right. .
\end{multline*}
The $\delta_{\phi}X^{abc\mid}{}_{d}$ part leaves $\cal{L}_{1}$ invariant modulo a total derivative. Moreover, it is easy to see that $\delta_{\phi}(\partial_{a}Z^{ab})=0$, and so it is straightforward that these transformations leave also invariant the last term in Lagrangian (\ref{eq:hordual}). The same is not invariant, however, under arbitrary shifts of $Z^{ab}$. Instead, demanding invariance of this term will restrict the arbitrariness of $\tilde{\omega}_{ab}$.

Let us define $Z^{0i}\equiv Z^{i}$ and separate it into transverse and longitudinal parts, $Z^{i}=Z_{\perp}^{i}+Z_{\parallel}^{i}$. We can do the same with the gauge parameters $\tilde{\omega}^{0i}\equiv\tilde{\omega}^{i}=\tilde{\omega}^{i}_\perp+\tilde{\omega}^{i}_\parallel$, the transverse (or divergenceless) $\tilde{\omega}^{i}_\perp$ corresponding to the shift of the component $Z_{\perp}^{i}$. Invariance of the last term in Lagrangian (\ref{eq:hordual}) requires 
\begin{equation*}
\delta_{\tilde{\omega}}(\partial_{i}Z^{i})=\delta_{\tilde{\omega}}(\partial_{i}Z_{\parallel}^{i})=\partial_{i}\tilde{\omega}^{i}=\partial_{i}\tilde{\omega}^{i}_\parallel=0,
\end{equation*}
an equation that puts no restriction on the gauge transformations of $Z_{\perp}^{i}$, which stand as pure gauge (as well as the components $Z^{ij}$). The parallel components $\tilde{\omega}^{i}_\parallel$ on the other hand are restricted to be constant in the spatial hypersurface, that is, $\delta_{\tilde{\omega}} Z_{\parallel}^{i}=\tilde{\omega}^{i}_\parallel(t)$.

We have here a very similar, if not the same, breaking of local Lorentz invariance as in the electric theory. To sum up, the gauge symmetries for the dual theory are given by
\begin{subequations}
\label{eq:dualgaugsymm} 
\begin{align}
&\delta_\phi X^{abc\mid}{}_{d}=\partial_{e}\phi^{abce\mid}{}_{d}+ \frac{3}{(d-2)}\delta_{d}^{[a}\partial_{e}\phi^{bc]e}, \label{eq:dgaugsymmh} \\
&\delta_{\tilde{\omega}}Z^{ij}=\tilde{\omega}^{ij}, \quad \delta_{\tilde{\omega}} Z^{i}=\tilde{\omega}^{i}_\perp+\tilde{\omega}^{i}_\parallel(t),~\text{such that}~\partial_{i}\tilde{\omega}^{i}_\perp=0. \label{eq:dgaugsymmf} 
\end{align}
\end{subequations}
We have constructed then the dual of Lagrangian (\ref{eq:lagrhorlin}) in $d$ dimensions, and identified the gauge symmetries of the same. We did it by giving a parent action defined with Lagrangian (\ref{eq:lagrhorpo}), from which we get both theories. In the case of the electric theory, we obtained linearized HL gravity from Lagrangian (\ref{eq:lagrhorlin}) by imposing the gauge fixing condition (\ref{eq:HLgaugefixcond}). We wonder now if there is an analogous magnetic gauge fixing condition for Lagrangian (\ref{eq:hordual}) in $d=4$ dimensions such that the projected dual theory in reduced space is linearized HL gravity. 

Let us introduce the field redefinition 
\begin{equation}
\label{eq:fieldred}
T_{a_1 \dots a_{d-3}\mid c}\equiv \frac{1}{3!}\epsilon_{a_1 \dots a_{d-3} efg}{X^{efg\mid}}_c,
\end{equation}
such that
\begin{equation*}
T_{[a_1 \dots a_{d-3}\mid c]}=0,
\end{equation*}
which makes manifest that the dual $\cal{L}_{1}[T,Z]$ is the FP Lagrangian for the dual graviton, in $d=4$ [see Eq. (2.16) in \cite{BoCnHe}]. In these variables, the action defined with Lagrangian (\ref{eq:hordual}) for the dual theory is written as
\begin{multline*}
    S_{\lambda}^{*}[T,Z]= \int\mathrm{d}^4x \Bigl( \cal{L}_{1}[T,Z]+\frac{4}{9}(\theta -1)(\partial_{i}Z^{i})^{2} \Bigr)\\
    =2 \int\mathrm{d}^4x\Bigl(\partial^aT^{bc}\partial_aT_{bc}-2\partial_aT^{ac}\partial^bT_{bc}+2\partial^bT^{a}{}_{a}\partial^c T_{bc}\\
    -\partial_aT^{b}{}_{b}\partial^a T^{c}{}_{c}  + \cal{M}[T,Z]
+ \frac{2}{9}(\theta -1)(\partial_{i} Z^{i})^{2}\Bigr).
\end{multline*}
In $d=4$ the symmetries (\ref{eq:dgaugsymmh}) can be written as $\phi^{abcd\mid}{}_{e}= \epsilon^{abcd}\chi_{e}$ with $\chi_{a}$ arbitrary, which implies $\delta_{\phi}T_{ab}=-\partial_{(a}\chi_{b)}$.

At this stage we can get rid of $Z^{ij}$ and $Z_{\perp}^{i}$ by fixing them to zero using the gauge invariance under arbitrary shifts. The $Z_{\parallel}^{i}$ component is, however, still part of the picture. In order to have a theory for $T_{ab}$ only, we need to impose a condition analogue to (\ref{eq:HLgaugefixcond}), with which we could write $Z_{\parallel}^{i}$ (or more precisely, its divergence) in terms of some components of $T_{ab}$.

Let us separate first the Lagrangian defining the action above in space and time components. That is,
\begin{multline*}
\cal{L}_{\lambda}[T,Z]=-2\Bigl( \dot{T}_{ij}\dot{T}^{ij}-\dot{T}^{2}-4\partial_{i}T_{0j}(\dot{T}^{ij}-\eta^{ij}\dot{T}) \\-2\partial^{i}T_{0i}\partial^{j}T_{0j}-2\partial_{i}T_{0j}\partial^{i}T^{0j}+2R_{T} \Bigr) \\
+\cal{M}[T,Z]+\frac{4}{9}(\theta -1)(\partial_{i} Z^{i})^{2},
\end{multline*}
where we have defined
\begin{multline*}
R_{T}\equiv -\frac12 \partial_{k}T_{ij}\partial^{k}T^{ij}+\frac12\partial_{i}T\partial^{i}T+\partial_{i}T^{ij}\partial^{k}T_{kj}\\
-\partial_{i}T^{ij}\partial_{j}T-T_{00}(\partial_{i}\partial_{j}T^{ij}-\Delta T)
\end{multline*}
[which is nothing but the potential (\ref{eq:grpot}) as a function of the component $T_{ij}$ of the dual graviton] and $T\equiv\eta^{ij}T_{ij}$.

With the following \textit{ansatz} for the magnetic projection condition,
\begin{equation}
\label{eq:magneticcond}
\rho^{*}\equiv \partial_{i}Z^{i}-(\dot{T}-2\partial^{i}T_{0i})=0,
\end{equation}
some arithmetics give
\begin{multline}
\label{eq:magHLd4}
\cal{L}_{\lambda}[T,Z]\mid_{\rho^{*}=0} ~\simeq~ -2\Bigl(\dot{T}_{ij}\dot{T}^{ij}-\textcolor{Blue}{\gamma} \dot{T}^2 -4\partial_iT_{0j}(\dot{T}^{ij} -\textcolor{Blue}{\gamma}\eta^{ij}\dot{T})\\+2(1-2\textcolor{Blue}{\gamma})\partial^iT_{0i}\partial^jT_{0j}-2\partial_iT_{0j}\partial^iT^{0j} +2R_{T}\Bigr),
\end{multline}
with
\begin{equation}
\label{eq:gamma}
\gamma\equiv 1+\frac29\Bigl(\frac{1-\lambda}{3\lambda-1}\Bigr).
\end{equation}
Here we were able to get rid of $\cal{M}$ because now this term is a total derivative that can be eliminated from the picture with boundary conditions on $T_{ab}$ (this is the only field which the above Lagrangian depends on). 

So, we see that it is possible to supplement $\cal{L}_\lambda[T,Z]$ with the condition (\ref{eq:magneticcond}) such that the reduced theory has the same structure as linearized HL gravity. The suggested \textit{ansatz} is motivated of course by recovering the linearized HL gravity Lagrangian,\footnote{Starting from a general combination of $\dot{T}$ and $\partial^{i}T_{0i}$ in it, which are the objects involved in the terms proportional to $\lambda$ in the Lagrangian for linearized HL gravity, it can be seen that those terms should be in a ratio of $\frac{\dot{T}}{\partial^{i}T_{0i}}\sim 2$.
} but there is another justification for this particular form of the gauge fixing condition. In the electric theory, we mentioned that the gauge fixing (\ref{eq:HLgaugefixcond}) projects the EoM for $e_i$ onto 
\begin{equation*}
\partial_j(\dot{h}-2\partial^i n_i)=\partial_j(2K)=0,    
\end{equation*}
with $K(x,t)$ the trace of the linearized extrinsic curvature. So, $\rho=0$ is implementing the constant mean curvature (CMC) gauge, in which $K$ is a function of time only.

In the case of the dual theory, the EoM for $Z^j$ is given by
\begin{equation*}
    \frac89 (\theta-1)\partial_j(\partial_i Z^i)=0,
\end{equation*}
from which we conclude that $\partial_i Z^i$ is a function of time only. So, the projection $\rho^*=0$ given by Eq. (\ref{eq:magneticcond}) is fixing this function to be the trace of the extrinsic curvature for the dual graviton, 
\begin{equation*}
    \partial_i Z^i = 2 K_T(t)=\dot{T}-2\partial^{i}T_{0i},
\end{equation*}
which is the CMC gauge again.

As in the electric theory, we want to take the dual Lagrangian (\ref{eq:hordual}) together with condition (\ref{eq:magneticcond}) as the dual theory of linearized HL gravity. Then, we would like to preserve such condition by the gauge symmetries of the theory. Applying transformations (\ref{eq:dualgaugsymm}) to $\rho^*=0$, where $\delta_{\phi}T_{ab}=-\partial_{(a}\chi_{b)}$, we get
\begin{align}
\label{eq:timerep}
0=\delta\rho^{*}&=\cancelto{0}{\delta_{\tilde{\omega}}(\partial_{i}Z^{i})}-\Bigl(\delta_{\phi}(\dot{T})-2\delta_{\phi}(\partial^{i}T_{0i})\Bigr) \notag \\
&=-\Bigl(\eta^{ij}\partial_{0}\delta_{\phi}T_{ij}-2\partial^{i}\delta_{\phi}T_{0i}\Bigr) \notag \\
&=-\Bigl( \eta^{ij}\partial_{0}(-\partial_{(i}\chi_{j)})-2\partial^{i}(-\partial_{(0}\chi_{i)}) \Bigr) \notag \\
\Rightarrow &-\Delta\chi_{0}=0.
\end{align}
This means that gauge symmetries such that $\chi_{0}=\chi_{0}(t)$ preserve the gauge fixing condition, while the same is invariant with $\chi_{i}$ completely arbitrary. These are the gauge symmetries of linearized HL gravity.

\begin{figure}[ht!]
\begin{center} 
\includegraphics[scale=0.45]{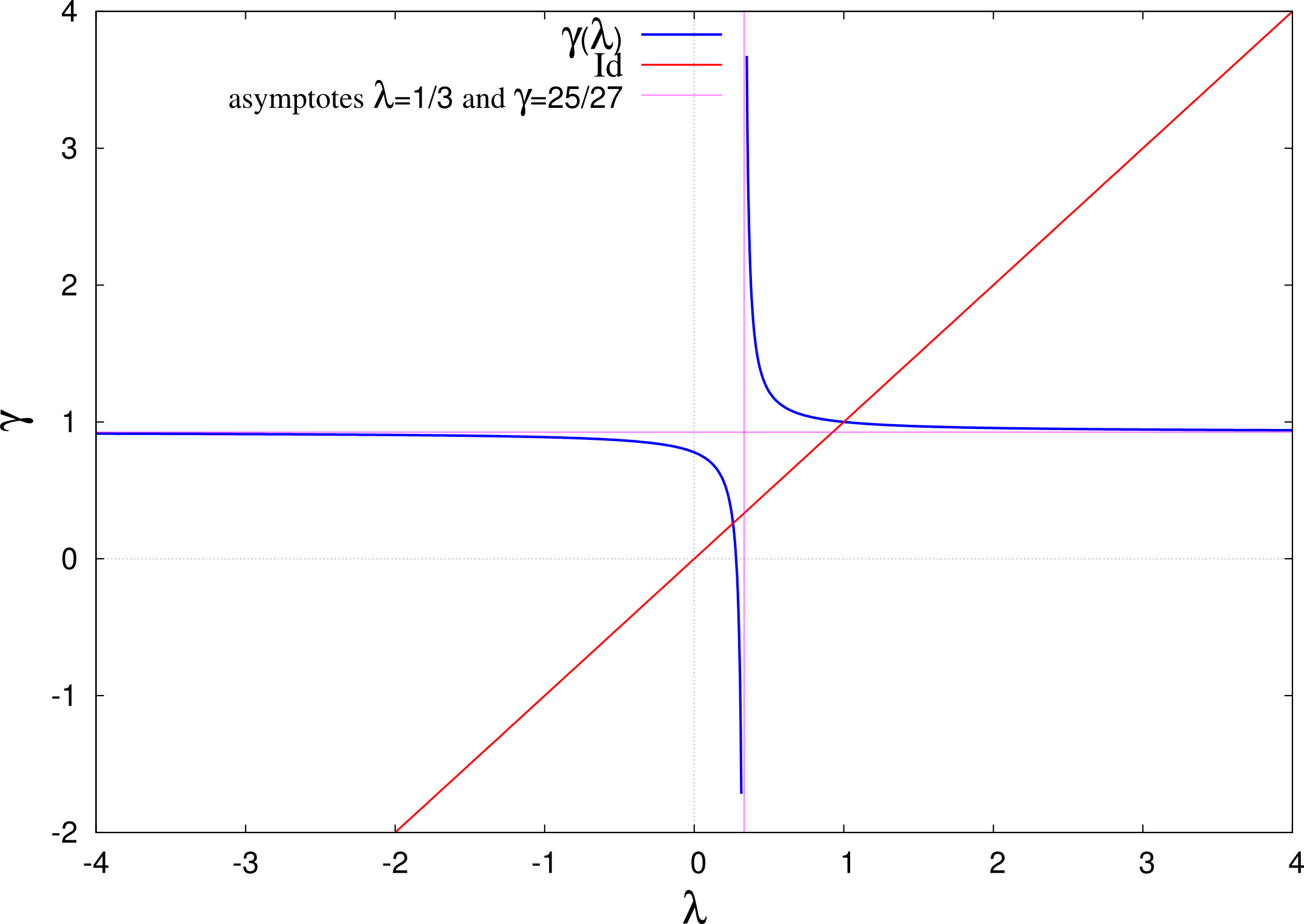} 
\end{center}
\caption{\textsf{\small{Plot of $\gamma$ \textit{vs.} $\lambda$.}}}\label{fig:gl}
\end{figure}

In conclusion, in $d=4$ the magnetic dual of linearized HL gravity Eq. (\ref{eq:magHLd4}) has the same structure of the former electric theory, but parametrized with $\gamma(\lambda)$. The electric and magnetic linearized HL deviations come then proportional to parameters with values in `complementary regimes'. For example, we see that $\frac13 <\lambda\leq 1 ~ \Rightarrow ~ \gamma \geq 1$. The case $\lambda=1=\gamma$ corresponds of course to FP theory on both sides of the duality, and the two parameters are equal once again for $\lambda=\frac{7}{27}=\gamma$ (see Fig.~\ref{fig:gl}).
For the latter, electric and magnetic linearized HL gravity are given by the same deviation from the FP Lagrangian with the same parameter. The two mentioned values of $\lambda$ belong to two branches in parameter space separated by an asymptote in $\lambda=\frac13$, where duality is not defined.

So far so good for $d=4$. What happens in $d>4$ dimensions? In the next section we analyze the particular case of $d=5$ and explore a HL-like deviation of the Curtright action for a mixed symmetry gauge field, the latter being the dual of the graviton in that case.

%%%----------------------------------------------------------------------------
\section{HL-like deviation of Curtright action\label{sec:Curtright}}
%%%----------------------------------------------------------------------------

The field redefinition (\ref{eq:fieldred}) in $d=5$ is $T_{ab|c}\equiv\frac{1}{3!}\epsilon_{abdef}X^{def|}{}_{c}$. The dual action defined with Lagrangian (\ref{eq:hordual}) in these variables is written
\begin{multline}
\label{eq:CurtrightallaHL}
    S_{\lambda}^{*}[T,Z]=\int\mathrm{d}^5x\cal{L}_{\lambda}[T,Z]=\\
     \int\mathrm{d}^5x \Bigl( \cal{L}_{\text{{\scriptsize C}}}[T]+ \cal{M}[T,Z]+\frac{2}{3}(\theta -1)(\partial_{i} Z^{i})^{2} \Bigr),
\end{multline}
where
\begin{multline*}
\cal{L}_{\text{{\scriptsize C}}}[T] = \partial_{a}T_{bc|d}\partial^{a}T^{bc|d}-\partial^{a}T_{bc|a}\partial_{d}T^{bc|d}
\\-2\partial_{a}T_{bc|}{}^{c}\partial^{a}T^{bd|}{}_{d} -4\partial^{a}T_{bc|a}\partial^{b}T^{cd|}{}_{d}\\
-2\partial_{a}T^{ac|d}\partial^{b}T_{bc|d}
+2\partial_{a}T^{ac|}{}_{c}\partial_{b}T^{bd|}{}_{d} 
\end{multline*}
is the Curtright Lagrangian \cite{Cu}. When $\lambda=\theta=1$, we know this Lagrangian is invariant under arbitrary shifts of the $Z^{ab}$ variables, and so the term $\cal{M}[T,Z]$ can be cast out. This is not the case for the action (\ref{eq:CurtrightallaHL}): Because of the form of the last term, $Z^{ij}$ and $Z^{i}_\perp$ components are still pure gauge, but not $Z^i_\parallel$. In the following we assume we have gotten rid of the pure gauge components. 

To take out of the picture the $Z^i$ variables and obtain a reduced theory just for the mixed symmetry $T_{ab|c}$, we impose a condition $\rho^{*}=\partial_i Z^i+\varrho[T]=0$ analogous to Eq. (\ref{eq:magneticcond}). So, by a HL-like deviation of the Curtright action we mean the above Lagrangian together with such a condition.

The function $\rho^{*}$ should be first order in derivatives (we want a second order reduced theory when we enforce $\rho^{*}=0$ into $\cal{L}_{\lambda}[T,Z]$) and involve $\eta_{ij}$ traces of the space and time components of $T_{ab|c}$. At the same time we expect that demanding preservation of such condition by the gauge transformations of the action (\ref{eq:CurtrightallaHL}) will restrict some of those symmetries. 

The gauge symmetries of $\cal{L}_{\text{{\scriptsize C}}}[T,Z]$, besides the shifts (\ref{eq:dgaugsymmf}), include the following transformations of the $T_{ab|c}$ fields:
\begin{equation}
\label{eq:symmCu}
\delta T_{ab|c}=\partial_{[a}S_{b]c}+\frac{2}{3}(\partial_{[a}A_{b]c}-\partial_{c}A_{ab}).
\end{equation}
Transformations given in Eq. (\ref{eq:dgaugsymmh}) can be recast in this form by realizing that the parameters can be written as $\phi^{abcd|}{}_{e}=-\epsilon^{abcdf}\chi_{fe}$ and using the separation $\chi_{ab}=S_{ab}+A_{ab}$, with $S_{ab}=S_{ba}$ and $A_{ab}=-A_{ba}$. The identity $T_{[ab|c]}=0$ has been implemented, and hence the third piece of Eq. (\ref{eq:symmCu}) (see \cite{BoCnHe,GaKn}). 

To build the function $\varrho[T]$ within $\rho^{*}=\partial_i Z^i+\varrho$, we first separate $\partial_{d}T_{ab|c}$ into space and time components, and then we observe that there are just three terms that can be used to form the traces of the desired form:
\begin{equation*}
\dot{T}_{0i|j},\qquad \partial_{i}T_{0j|0},\qquad\text{and}\qquad \partial_{i}T_{jk|l}.
\end{equation*}
These are
\begin{equation*}
\dot{T}_{0},\qquad \partial^{i}T_{0i|0},\qquad\text{and}\qquad \partial^{i}T_{i},
\end{equation*}
where $T_{a}\equiv T_{a i|}{}^i=T_{a i|j}\eta^{ij}$ (analogous to $T=T_i{}^i$, the trace of $T_{ij}$ in $d=4$). Finally, we define $\varrho$ as the general linear combination
\begin{equation}
\label{eq:condmagCu}
\varrho=\upsilon_{1}\dot{T}_{0}+\upsilon_{2}\partial^{i}T_{i}+\upsilon_{3}\partial^{i}T_{0i|0}.
\end{equation}

The variation of $\rho^{*}=0$ gives
\begin{align*}
0=\delta\rho^{*}=&\cancelto{0}{\delta(\partial_{i}Z^{i})} \\
&+\frac12\Bigl(\upsilon_{1}\ddot{S}+\upsilon_{2}(\eta^{ij}\Delta-\partial^{i}\partial^{j})S_{ij}\Bigr) -\frac12\upsilon_{3}\Delta S_{00} \\
&+\frac12(\upsilon_{3}-\upsilon_{1})\partial^{i}\dot{S}_{0i}-(\upsilon_{3}+\upsilon_{1})\partial^{i}\dot{A}_{0i}.
\end{align*}
Here $S\equiv\eta^{ij}S_{ij}$. This is definitely more complicated than Eq.  (\ref{eq:timerep}) for the $d=4$ case, but we immediately see that the choice $\upsilon_{3}=-\upsilon_{1}$ makes $\rho^{*}$ invariant under the gauge transformations with parameters $A_{ab}$, and poses a differential equation on the parameters $S_{ab}$. This differential equation can be written
\begin{equation*}
2\upsilon_{1}\eta^{ij}\partial_{[0}S_{i][j,0]} -\frac{\upsilon_{2}}{2}(\partial^{i}\partial^{j}S_{ij}-\Delta S)=0
\end{equation*}
(the ``$,$'' at the end of the first term means partial derivative).

The combination given in Eq. (\ref{eq:condmagCu}) is not unfamiliar: For $\upsilon_1=1$, $\upsilon_2=0$, and $\upsilon_3=-2$, $\varrho=0$ matches with a component of the gauge fixing condition that is imposed on the Lagrangian EoM coming from $\cal{L}_{\text{{\scriptsize C}}}[T]$ to get the wave equation $\square T_{ab|c}=0$\footnote{In other words, this choice would be the analogue of the transverse-traceless gauge.} [see eq. (16) in \cite{Cu}]. The choice $\upsilon_{3}=-\upsilon_{1}$ is a different gauge, and will deliver a nonrelativistic wave equation.

We will not go further in the analysis of this deviation of the Curtright action in the present work. A proper study of nonrelativistic deviations of this theory deserves its own space.

%%%----------------------------------------------------------------------------
\section{Summary and conclusions\label{sec:conclu}}
%%%----------------------------------------------------------------------------

In this work the powerful tool of parental actions to construct duality relations in classical field theories has been implemented in a nonrelativistic theory of gravity. We followed the scheme of the authors of \cite{BoCnHe}, and hence we proceeded in two steps. First we constructed a covariant action in the \textit{vielbein} basis defined with the Lagrangian $\cal{L}_\lambda[e_{ab}]$, which entails a deviation from FP theory in the framelike formalism [see Lagrangian (\ref{eq:lagrhorlin})] and that in an appropriate gauge reduces to linearized HL gravity. The gauge fixing (\ref{eq:HLgaugefixcond}) is such that it breaks the diffeomorphism invariance to a subgroup, precisely the group of foliation preserving diffeomorphisms of HL gravity (in the linear approximation). In a second step we found a parent action defined with Lagrangian (\ref{eq:lagrhorpo}), from which both $\cal{L}_\lambda[e_{ab}]$ and its dual partner Eq. (\ref{eq:hordual}) can be derived. The further gauge fixing (\ref{eq:magneticcond}) on the dual branch left us with the magnetic dual of linearized HL gravity. We observed that the gauge fixing in both sides of the duality was just realizations of the CMC gauge. 

We found that the magnetic dual of linearized HL gravity in $d=4$ has the same structure as the electric theory, and moreover that for the value $\lambda=\frac{7}{27}$ of the characteristic parameter, the theory is self-dual. This example shows then that self-duality is not tied to Lorentz invariance. In the case of $d=5$, we suggested a nonrelativistic deviation of the Curtright action by imposing a gauge fixing condition analogous to the one used in the case of $d=4$. Breaking Lorentz invariance at the linearized level in this way might unlock nonlinear extensions \textit{á la} Ho\v{r}ava of the Curtright action. A different approach for such a nonrelativistic deformation could arise from studying a deviation similar to Lagrangian (\ref{eq:lagrhorpo}) but in the parent action proposed by the authors of \cite{BoHo} to implement duality in nonlinear Einstein theory.

Finally, we want to make an observation based on the result reported here for the dual of linearized HL gravity with characteristic parameter $\lambda$. In $d=4$ the magnetic dual theory for the dual graviton was found to have a Lagrangian with the same structure as the electric linearized HL gravity for the graviton. Its characteristic parameter $\gamma$ given in (\ref{eq:gamma}), however, is in general different from $\lambda$. With the theory for the dual graviton having the same (tensorial) structure as the theory for the graviton, we can take the magnetic theory as input in the bottom of the left branch of the diagram depicted in Fig. \ref{fig:approach}, \textit{i.e.}, as an electric theory, to perform a sort of iteration process.

The output of this iteration will be a theory with the same structure as linearized HL gravity again, but the characteristic parameter will be given by  
\begin{equation}
\label{eq:gammap}
\gamma'\equiv 1+\frac29\Bigl(\frac{1-\gamma(\lambda)}{3\gamma(\lambda)-1}\Bigr),
\end{equation}
with $\gamma'\neq\lambda$. This iteration process is of course only possible in $d=4$, where the magnetic theory (the output) possesses the same tensorial structure as the original electric theory (the input). It is clear that in $d=5$, for example, this iteration exercise cannot be performed. 

The same kind of iteration exercise can be thought of in the relativistic $\lambda=1$ case. There, however, the output not only possesses the same structure but also the same value of the characteristic parameter. The same happens with the nonrelativistic theory with $\lambda=\frac{7}{27}$. These two values behave then like ``equilibrium points'' of the iteration. Hence, in $d=4$ and after any number of iterations when $\lambda\neq 1,\frac{7}{27}$, the output will be always linearized HL gravity with a different parameter each time.
\begin{figure}[ht!]
\begin{center}   
\includegraphics[scale=0.45]{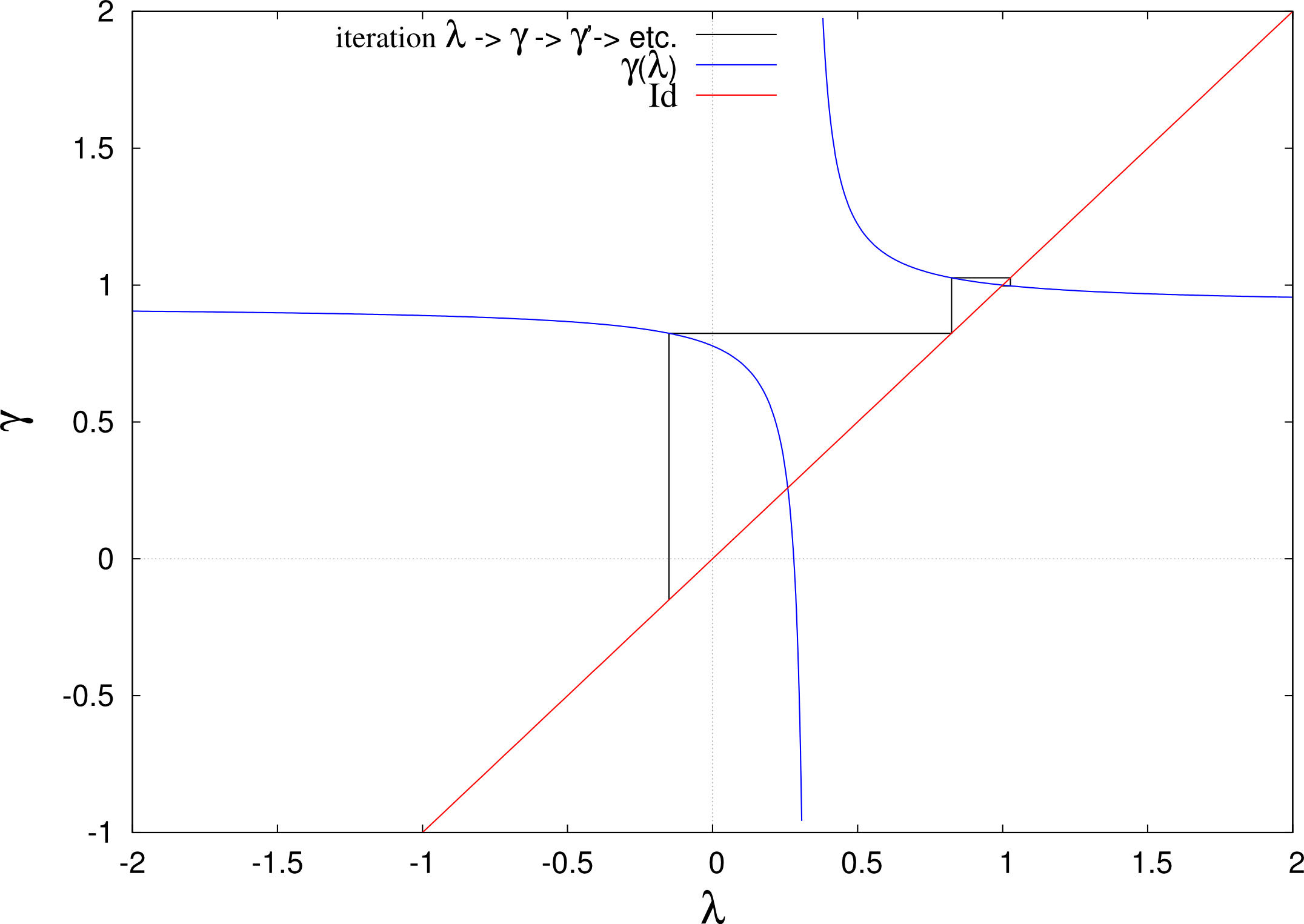}
\end{center}
\caption{\textsf{\small{Sample of three iterations starting in $\lambda=-0.15$.}}}\label{fig:g11}
\end{figure}

The peculiarity that we want to mention is that the parameters resulting after each iteration come closer and closer to 1. This value is a sort of `stable equilibrium point', while $\lambda=\frac{7}{27}$ is an `unstable equilibrium point', because in contrast to the former, the values of the parameters resulting from successive iterations run away from it. In Fig.~\ref{fig:g11} a picture of the generic behavior is presented. The same results no matter what the initial value of $\lambda$ is, with the exceptions of the equilibrium points, and of any value that after a finite number of dualizations gives $\frac{1}{3}$, for which the duality map is ill defined. We do not know what the content of this curious behavior is, the study of which we leave for future work.

The nonrelativistic model we used for this exploration was a linearized Lagrangian in flat spacetime. It would be interesting to perform the same analysis within (a)dS or Lifshitz backgrounds. This could be of interest, perhaps, for recent discussions regarding the nonrelativistic holographic correspondence \cite{JaKa,GrHoMe-Th}.

%%%%%%%%%%%%%%%%%%%%%%%%%%%%%%%%%%%%%%%%%%%%%%%%%%%%%%%%%%%%%%%%%%%%%%%%%%%%%%%
%%%%%%%%%%%%%%%%%%%%%%%%%%%%%%%%%%%%%%%%%%%%%%%%%%%%%%%%%%%%%%%%%%%%%%%%%%%%%%%

\section*{Acknowledgements}

I.C. thanks Riccardo Argurio, Glenn Barnich, and Gustavo Lucena Gómez for useful discussions during the elaboration of this work. I.C. also thanks Nicolas Boulanger and Dmitry Ponomarev for discussions and comments on the results. This work is partially supported by SEP-CONACyT CB-2008-104649 grant (Mexico) \textit{Aplicaciones de la teor\'ia de cuerdas al plasma de quarks y gluones}. The work of I.C. is partially supported by CONACyT (Mexico) through the \textit{Estancias Sab\'aticas y Posdoctorales al Extranjero para la Consolidaci\'on de Grupos de Investigaci\'on} program, by IISN - Belgium (conventions 4.4511.06 and 4.4514.08), by the \textit{Communaut\'e Fran\c{c}aise de Belgique} through the ARC program and by the ERC through the ``SyDuGraM'' Advanced Grant.

%%%%%%%%%%%%%%%%%%%%%%%%%%%%%%%%%%%%%%%%%%%%%%%%%%%%%%%%%%%%%%%%%%%%%%%%%%%%%%%
%%%%%%%%%%%%%%%%%%%%%%%%%%%%%%%%%%%%%%%%%%%%%%%%%%%%%%%%%%%%%%%%%%%%%%%%%%%%%%%

\appendix

%%%%----------------------------------------------------------------------------
\section*{Appendix: Constraint structure of $\cal{L}_{\lambda}[e]$\label{app:newconstr}}
%%%%----------------------------------------------------------------------------

In this section we show that $\rho=0$ in Eq. (\ref{eq:HLgaugefixcond}) is a gauge fixing condition for a first class constraint of Lagrangian $\cal{L}_{\lambda}[h_{ij},n_{i},n;e_{i}]$. We will just summarize the highlights of the constraint analysis for it. We assume here we already eliminated $f_{ij}$ and the transverse part of $e_{i}$ according to what has been described in Sec. \ref{sec:linHLvierbein} ($e_{i}$ denotes in what follows the longitudinal part of this field).

Since $\cal{L}_{\lambda}$ does not depend on time derivatives of $n,n_{i},e_{i}$, we have the primary constraints $\phi^{0}=p$, $\phi_{n}^{i}=p^{i}$ and $\phi_{e}^{i}=q^{i}$ for the momenta conjugated to those variables. The canonical Hamiltonian is given by
\begin{equation*}
    H=\int\mathrm{d}^3x \Bigl[\frac12 \pi^{ij}\pi_{ij} -\frac{\theta}{4}\pi^2+2\pi^{ij}\partial_{i}n_{j}-R+(\theta-1)(\pi-\rho)\rho\Bigr],
\end{equation*}
where $\theta$, $\rho$, and $R$ are the same as in Sec. \ref{sec:diracanalisis}. Stabilization of the primary constraints delivers a set of secondary constraints given by
\begin{equation*}
    \psi^0=2(\partial_i\partial_j h^{ij}-\Delta h), \quad \psi_{n}^i=2\partial_j \pi^{ji}+(\theta-1)\partial^{i}(\pi-2\rho), 
\end{equation*}
and
\begin{equation*}
\psi_{e}^{i}=(1-\theta)\partial^i(\pi-2\rho).
\end{equation*}
In contrast to the Dirac analysis carried out for $\cal{L}_{\text{{\scriptsize HL}}}$, there are no more secondary constraints coming from stabilization of these, but a condition on Lagrange multipliers. 

From the Poisson brackets between all these constraints we get at first sight a very odd result. There is just one first class constraint, $\phi^{0}$, and the rest (13 of them) are second class. This is odd because we do not see the first class constraints generating the gauge transformations (\ref{eq:gaugsymm}), particularly the diffeomorphisms in the spatial hypersurfaces. Besides, the amount of second class constraints being odd gives a noninteger number for the d.o.f. count.

We can, however, make a redefinition of the set of constraints that reveals the first class content that generates the gauge transformations of the theory. This is
\begin{equation}\label{eq:constrredef}
\left.
\begin{aligned}
\phi^{0},&\phi_{n}^{i}, \phi_{e}^{i}\\
\psi^{0},&\psi_{n}^{i}, \psi_{e}^{i}
\end{aligned}
\right\}
\sim\left\{
\begin{aligned}
&\phi^{0},~\phi_{+}^{i}\equiv\frac12(\phi_{n}^{i}+\phi_{e}^{i}),~\phi_{-}^{i}\equiv\frac12(\phi_{n}^{i}-\phi_{e}^{i})\\
&\tilde{\psi}^{0}\equiv\frac12\psi^{0}-\partial_{i}\phi_{-}^{i},\tilde{\psi}^{i}\equiv\psi_{n}^{i}+ \psi_{e}^{i}, \psi_{e}^{i}
\end{aligned}
\right. 
\end{equation}
(notice that $\tilde{\psi}^i=\psi^{i}=2\partial_{j}\pi^{ji}$, with $\psi^{i}$ the momentum constraint of Sec. \ref{sec:diracanalisis}), which at the end delivers the following classification:
\begin{center}
\begin{tabular}{lr}

    first class & $\phi^0$, $\phi_{+}^i$, $\tilde{\psi}^{0}$, $\tilde{\psi}^i$  \\ 
    second class & $\phi_{-}^i$, $\psi_{e}^{i}$ \\
    
\end{tabular}.
\end{center}

The gauge transformations generated by the set of first class constraints are given by
\begin{eqnarray*}
\phi^{0}&:&\delta_{u} n = u, \\
\phi_{+}^{i}&:& \delta_{u_{+}} n_{i}=\delta_{u_{+}} e_{i}=\frac12 u_{+}{}_{i}, \\
\tilde{\psi}^{0}&:&\delta_{v}\pi^{ij}=\eta^{ij}\Delta v-\partial^{i}\partial^{j}v,~\delta_{v}n_{i}=-\delta_{v}e_{i}=\frac12\partial_{i}v, \\
\tilde{\psi}^{i}&:& \delta_{w}h_{ij}=-2\partial_{(i}w_{j)},
\end{eqnarray*}
for arbitrary $u,u_{+}{}_{i},v,w_{i}$, functions of $x$ and $t$. The gauge transformations (\ref{eq:gaugsymm}) are reproduced by these for the parameters $w_{i}=-\xi_{i},~v=2\xi_{0},~u_{+}{}_{i}=2\dot{\xi}_{i},~u=-\dot{\xi}_{0}$. As can be checked directly, the counting of d.o.f. gives 2.

Now we can discern the role of the condition $\rho=0$ defined in Sec. \ref{sec:linHLvierbein}. This condition is a gauge fixing constraint for the first class $\tilde{\psi}^{0}$, forming together a second class pair:
\begin{equation*}
\{\rho(x),\tilde{\psi}^{0}(y)\}=-\Delta\delta(x-y).
\end{equation*}
This constraint is accessible: If we have $\partial^{i}n_{i}-\partial^{i}e_{i}\equiv m(x)\neq 0$, a gauge transformation $n'_i\equiv n_i +\delta_v n_i$ and $e'_i\equiv e_i +\delta_v e_i$ gives
\begin{align*}
    \partial^i n'_i-\partial^i e'_i&=\partial^i n_i-\partial^i e_i +\partial^i(\delta_v n_i-\delta_v e_i)\\
    & = m -\Delta v.
\end{align*}
There exists $v(x)$ such that $m-\Delta v=0$, and it is 
$$
v=-\int \mathrm{d}^3 y \nabla\cdot \Bigl(\frac{(\vec{n}-\vec{e})(x)}{4\pi\mid\vec{x}-\vec{y}\mid}\Bigr),
$$
as can be easily deduced noticing that $m(x)$ is the divergence of the longitudinal component of $\vec{n}-\vec{e}$:
$$
m(x)=\nabla\cdot(\vec{n}-\vec{e})=\nabla\cdot(\vec{n}_\parallel-\vec{e}_\parallel),
$$
and $\vec{n}_\parallel-\vec{e}_\parallel=\nabla\Bigl[-\int \mathrm{d}^3 y \nabla\cdot \Bigl(\frac{(\vec{n}-\vec{e})(x)}{4\pi\mid\vec{x}-\vec{y}\mid}\Bigr)\Bigr]$.
In conclusion, $\rho=0$ can be part of a full set of gauge fixing constraints.

%%%%%%%%%%%%%%%%%%%%%%%%%%%%%%%%%%%%%%%%%%%%%%%%%%%%%%%%%%%%%%%%%%%%%%%%%%%%%%%
%%%%%%%%%%%%%%%%%%%%%%%%%%%%%%%%%%%%%%%%%%%%%%%%%%%%%%%%%%%%%%%%%%%%%%%%%%%%%%%

\end{document}